\documentclass[12pt,a4paper]{article}
\usepackage{amsfonts}
\usepackage{amssymb}
\usepackage{latexsym}
\begin{document}

\begin{center}
{\Large \bf Submanifolds in space-time with unphysical extra
dimensions, cosmology and warped brane world models} \\

\vspace{4mm}

Mikhail N. Smolyakov\\ \vspace{0.5cm} Skobeltsyn Institute of
Nuclear Physics, Moscow State University
\\ 119991
Moscow, Russia\\
\end{center}

\begin{abstract}
The explicit coordinate transformations which show the equivalence
between a four-dimensional spatially flat cosmology and an
appropriate submanifold in the flat five-dimensional Minkowski
space-time are presented. Analogous procedure is made for the case
of five-dimensional warped brane world models. Several examples
are presented.
\end{abstract}

It is well known that in the general case a four-dimensional
pseudo-Riemannian manifold can be represented as a submanifold in
a flat ten-dimensional space-time \cite{Eddington}. In the case of
additional symmetries the dimensionality of the ambient space-time
may be smaller. The well known example is the de Sitter space
$dS_{4}$, which can be represented as a hyperboloid in the
five-dimensional Minkowski space-time \cite{Weinberg,Misner}.
Below we will show explicitly that any four-dimensional space-time
corresponding to the spatially-flat cosmology can be defined as a
submanifold in the five-dimensional Minkowski space-time, whereas
space-time corresponding to some five-dimensional warped brane
world models can be defined as a submanifold in the
six-dimensional flat space-time.

First, let us consider a five-dimensional space with the flat
metric
\begin{equation}\label{metric}
ds^{2}=-dt'^{2}+d{\vec y}^{2}+dz^{2}.
\end{equation}
Let us represent the coordinates in the following form
\begin{eqnarray}\label{subst1}
t'&=&\frac{1}{\alpha}\left(a(t)\vec x^{2}+\int\frac{dt}{\dot a(t)}\right)+\alpha\frac{a(t)}{4},\\
\label{subst2} z&=&\frac{1}{\alpha}\left(a(t)\vec
x^{2}+\int\frac{dt}{\dot a(t)}\right)-\alpha\frac{a(t)}{4},\\
\label{subst3} \vec y&=&a(t)\vec x,
\end{eqnarray}
where $\dot a(t)=\frac{da(t)}{dt}$, $\alpha$ is a constant with
the dimension of length, $\alpha\neq 0$. Substituting
(\ref{subst1})-(\ref{subst3}) into (\ref{metric}) we easily obtain
\begin{equation}\label{metric1}
ds^{2}=-dt^{2}+a^{2}(t)d{\vec x}^{2},
\end{equation}
which corresponds to a cosmology with zero spatial curvature.

Now let us find an explicit form of the appropriate manifold in
accordance with the parametric coordinate representation
(\ref{subst1}), (\ref{subst2}) and (\ref{subst3}). From
(\ref{subst1}) and (\ref{subst2}) we get
\begin{equation}\label{time}
2(t'-z)=\alpha a(t).
\end{equation}
Adding (\ref{subst1}) to (\ref{subst2}), multiplying the resulting
sum by (\ref{time}) and using (\ref{subst3}), we obtain
\begin{equation}\label{manifold}
t'^{2}-\vec
y^{2}-z^{2}=\left.\frac{2(t'-z)}{\alpha}\int\frac{dt}{\dot
a(t)}\right|_{t=a^{-1}\left(\frac{2(t'-z)}{\alpha}\right)},
\end{equation}
which can be rewritten as
\begin{equation}\label{manifold1}
t'^{2}-\vec
y^{2}-z^{2}=\left[a\int\frac{da}{a^{2}H^{2}(a)}\right]_{a=\frac{2(t'-z)}{\alpha}},
\end{equation}
where $H(a)=\frac{\dot a(t)}{a(t)}$ is the Hubble parameter.
Equation (\ref{manifold1}) describes a four-dimensional
submanifold, embedded into the flat five-dimensional space-time
with coordinates $t',\vec y,z$, for the general form of $a(t)$.

For simplicity we omit the integration constant appearing in
$\int\frac{da}{a^{2}H^{2}(a)}$. Indeed, the term
$$c\,a|_{a=\frac{2(t'-z)}{\alpha}}=\frac{2c(t'-z)}{\alpha}$$ in (\ref{manifold1}),
where $c$ is an integration constant, can be eliminated by the
coordinate transformation $t'\to t'+\frac{c}{\alpha}$, $z\to
z+\frac{c}{\alpha}$. Now let us turn to specific examples.
\begin{itemize}
\item Radiation dominated Universe (equation of state parameter
$\omega=1/3$):
$$a\sim \sqrt{\lambda t},$$
$$
t'^{2}-\vec
y^{2}-z^{2}=\frac{64}{3\alpha^{4}\lambda^{2}}\left(t'-z\right)^{4}.
$$
\item Matter dominated Universe ($\omega=0$):
$$a\sim (\lambda t)^{2/3},$$
$$
t'^{2}-\vec
y^{2}-z^{2}=\frac{9}{\alpha^{3}\lambda^{2}}\left(t'-z\right)^{3}.
$$
\item $\omega=-1/3$:
$$a\sim \lambda t,\qquad (\ddot a(t)=0),$$
$$
t'^{2}-\vec
y^{2}-z^{2}=\frac{4}{\alpha^{2}\lambda^{2}}\left(t'-z\right)^{2}.
$$
Note, that in this case there exists the dilatation symmetry of
the manifold $t',z,\vec y\to \beta t',\beta z,\beta\vec y$, where
$\beta$ is a constant. If $\alpha=\frac{2}{\lambda}$, we get
$$
t'z=\frac{\vec y^{2}}{2}+z^{2},
$$
which is linear in $t'$.

\item $\omega=-2/3$:
$$a\sim (\lambda t)^{2},$$
$$
t'^{2}-\vec
y^{2}-z^{2}=\frac{(t'-z)}{2\alpha\lambda^{2}}\ln\left(\frac{2(t'-z)}{\alpha}\right).
$$

\item Cosmological constant ($\omega=-1$):
$$a\sim e^{\lambda t},$$
$$
t'^{2}-\vec y^{2}-z^{2}=-\frac{1}{\lambda^{2}}.
$$
This result is well known and can be found in
\cite{Weinberg,Misner}. If one takes $\alpha=\frac{2}{\lambda}$,
coordinate transformations (\ref{subst1})--(\ref{subst3}) take the
form \cite{Weinberg,Misner}:
\begin{eqnarray}
t'&=&\frac{\lambda}{2}e^{\lambda t}\vec x^{2}+\frac{1}{\lambda}sh(\lambda t),\\
z&=&\frac{\lambda}{2}e^{\lambda t}\vec x^{2}-\frac{1}{\lambda}ch(\lambda t),\\
\vec y&=&e^{\lambda t}\vec x.
\end{eqnarray}

\item General case ($\omega\neq -1$, $\omega\neq -2/3$):
$$a\sim (\lambda t)^{\frac{2}{3+3\omega}},$$
see, for example, \cite{Rubakov:2005tx}, and
$$
t'^{2}-\vec
y^{2}-z^{2}=\frac{9(1+\omega)^{2}}{4\lambda^{2}(2+3\omega)}\left(\frac{2(t'-z)}{\alpha}\right)^{3(1+\omega)}.
$$
\end{itemize}

Now let us turn to five-dimensional brane world models with flat
four-dimensional metric on the brane. Let us consider the
six-dimensional space-time with the metric
\begin{equation}\label{metric-brane}
ds^{2}=\eta_{\mu\nu}dX^{\mu}dX^{\nu}+dY^{2}-dZ^{2},
\end{equation}
where $\mu,\nu=0,1,2,3$, $\eta_{\mu\nu}=diag(-1,1,1,1)$. Making
substitution
\begin{eqnarray}\label{subst-b1}
Y&=&\frac{1}{\alpha}\left(A(y)\eta_{\rho\sigma}x^{\rho}x^{\sigma}-\int\frac{dy}{dA/dy}\right)-\alpha\frac{A(y)}{4},\\
\label{subst-b2}
Z&=&\frac{1}{\alpha}\left(A(y)\eta_{\rho\sigma}x^{\rho}x^{\sigma}-\int\frac{dy}{dA/dy}\right)+\alpha\frac{A(y)}{4},\\
\label{subst-b3} X^{\mu}&=&A(y)x^{\mu},
\end{eqnarray}
where $\alpha$ is a constant, into (\ref{metric-brane}), we get
\begin{equation}\label{metric-brane1}
ds^{2}=A^{2}(y)\eta_{\mu\nu}dx^{\mu}dx^{\nu}+dy^{2},
\end{equation}
which is the standard form of the metric in such models (see
\cite{Rubakov:2001kp,Kubyshin:2001mc}). The corresponding
submanifold in the six-dimensional flat space-time can be obtained
in a way analogous to that presented above and takes the form
\begin{equation}\label{manifold-brane}
-\eta_{\mu\nu}X^{\mu}X^{\nu}-Y^{2}+Z^{2}=\left[-A(y)\int\frac{dy}{dA/dy}\right]_{y=A^{-1}\left(\frac{2(Z-Y)}{\alpha}\right)}.
\end{equation}

As an example, let us consider the simplest case $A=e^{-k|y|}$,
discussed in papers
\cite{Randall:1999ee,Randall:1999vf,Grzadkowski:2003fx,Mikhailov:2006vx}.
For simplicity we also take $e^{-k|y|}\to e^{-ky}$. From
(\ref{manifold-brane}) we obtain
\begin{equation}\label{manifold-RS}
-\eta_{\mu\nu}X^{\mu}X^{\nu}-Y^{2}+Z^{2}=\frac{1}{k^{2}}.
\end{equation}
One can easily see that this submanifold in six-dimensional flat
space-time with metric (\ref{metric-brane}) corresponds to the
anti-de Sitter space-time $AdS_{5}$, which is of course the well
known result (indeed, (\ref{metric-brane1}) with $A=e^{-ky}$ is
simply the metric of $AdS_{5}$ in horospherical coordinates). With
$\alpha=\frac{2}{k}$ coordinate transformations
(\ref{subst-b1})--(\ref{subst-b3}) take the form
\begin{eqnarray}
Y&=&\frac{k}{2}e^{-ky}\eta_{\rho\sigma}x^{\rho}x^{\sigma}+\frac{1}{k}sh(ky),\\
Z&=&\frac{k}{2}e^{-ky}\eta_{\rho\sigma}x^{\rho}x^{\sigma}+\frac{1}{k}ch(ky),\\
X^{\mu}&=&e^{-ky}x^{\mu}.
\end{eqnarray}

We hope that the results presented in this note can be interesting
from theoretical and pedagogical points of view.

\subsection*{Acknowledgments}
The author is grateful to G.Yu.~Bogoslovsky and I.P.~Volobuev for
valuable discussions. The work was supported by grant of Russian
Ministry of Education and Science NS-1456.2008.2, grant for young
scientists MK-5602.2008.2 of the President of Russian Federation,
grant of the "Dynasty" Foundation and scholarship for young
scientists of Skobeltsyn Institute of Nuclear Physics of
M.V.~Lomonosov Moscow State University.

\end{document}